\documentclass[12pt]{article}

\usepackage{amssymb}
\usepackage{amsmath}
\usepackage{amsfonts}

\newcommand{\R}{\mathbb{R}}
\newcommand{\C}{\mathbb{C}}
\newcommand{\Z}{\mathbb{Z}}

\newcommand{\be}{\begin{equation}}
\newcommand{\ee}{\end{equation}}
\newcommand{\bea}{\begin{eqnarray}}
\newcommand{\eea}{\end{eqnarray}}
\newcommand{\nn}{\nonumber}
\newcommand{\kt}{\rangle}
\newcommand{\br}{\langle}

\newcommand{\ed}{\end{document}}
\newcommand{\bbr}{\br\!\br}
\newcommand{\kkt}{\kt\!\kt}

\newcommand{\dto}{\Leftrightarrow}

\begin{document}

\title{Pseudo-Unitary Operators and Pseudo-Unitary Quantum~Dynamics}
\author{Ali Mostafazadeh\thanks{E-mail address: amostafazadeh@ku.edu.tr}\\ \\
Department of Mathematics, Ko\c{c} University,\\
Rumelifeneri Yolu, 34450 Sariyer, Istanbul, Turkey}
\date{ }
\maketitle

\begin{abstract}
We consider pseudo-unitary quantum systems and discuss various
properties of pseudo-unitary operators. In particular we prove a
characterization theorem for block-diagonalizable pseudo-unitary
operators with finite-dimensional diagonal blocks. Furthermore, we
show that every pseudo-unitary matrix is the exponential of
$i=\sqrt{-1}$ times a pseudo-Hermitian matrix, and determine the
structure of the Lie groups consisting of pseudo-unitary matrices.
In particular, we present a thorough treatment of $2\times 2$
pseudo-unitary matrices and discuss an example of a quantum system
with a $2\times 2$ pseudo-unitary dynamical group. As other
applications of our general results we give a proof of the
spectral theorem for symplectic transformations of classical
mechanics, demonstrate the coincidence of the symplectic group
$Sp(2n)$ with the real subgroup of a matrix group that is
isomorphic to the pseudo-unitary group $U(n,n)$, and elaborate on
an approach to second quantization that makes use of the
underlying pseudo-unitary dynamical groups.
\end{abstract}

\baselineskip=24pt

\section{Introduction}

Past two years have witnessed a growing interest in
pseudo-Hermitian Hamiltonians, \cite{p1} -- \cite{za-03}.
Initially, the concept of a pseudo-Hermitian operator was
developed to describe the mathematical structure of (the possibly
non-unitary) $PT$-symmetric quantum systems, \cite{p1,p2}. Then it
became clear that any diagonalizable Hamiltonian that admitted a
symmetry generated by an invertible antilinear operator was
necessarily pseudo-Hermitian, \cite{p3,ss1}. The intriguing
spectral properties of pseudo-Hermitian Hamiltonians generalize to
the class of block-diagonalizable Hamiltonians with
finite-dimensional blocks \cite{p6}, so does the connection with
antilinear symmetries \cite{ss2}. Among the most important
outcomes of the study of pseudo-Hermitian Hamiltonians is the
recent solution of the old problem of constructing invariant
positive-definite inner products on the solution space of the
Klein-Gordon-type equations, \cite{cqg,ap}.

A quantum system with a (time-independent) pseudo-Hermitian
Hamiltonian has necessarily a pseudo-unitary evolution.
Pseudo-unitary quantum systems with a two-dimensional Hilbert
space provide the simplest nontrivial examples of such systems. As
shown in Ref.~\cite{cqg}, a classical simple harmonic oscillator
is equivalent to a pseudo-unitary quantum system with a
two-dimensional Hilbert space. Recently Ahmed and Jain
\cite{aj,aj-03} and Ahmed \cite{za-03} have considered the
application of certain $2\times 2$ pseudo-Hermitian matrices in
statistical mechanics and elaborated on the fact that they form a
Lie algebra.

The purpose of this article is three fold. First, we use the
method of Ref.~\cite{p6} to obtain a characterization of the
block-diagonalizable pseudo-unitary operators having
finite-dimensional diagonal blocks. Next, we confine our attention
to pseudo-unitary matrices and show that they are obtained by
exponentiating pseudo-Hermitian matrices. This is a nontrivial
result, because, for a fixed $\eta$, not every
$\eta$-pseudo-unitary matrix is the exponential of $i=\sqrt{-1}$
times an $\eta$-pseudo-Hermitian matrix. Finally, we emphasize
that unlike the set of $\eta$-pseudo-unitary operators (with a
fixed $\eta$), the set of all pseudo-unitary operators does not
form a group. If the Hilbert space in which these operators act is
finite-dimensional, then the group of $\eta$-pseudo-unitary
operators is isomorphic to one of the groups $U(n)$ or $U(n,m)$
for some $m,n\in\Z^+$. For example, the Lie algebra of the
pseudo-unitary matrices constructed in Ref.~\cite{aj} is
isomorphic to $u(1,1)$. This follows from the fact that the
corresponding inner product is indefinite; there is no need to go
through the calculation of the structure constants as done in
Ref.~\cite{aj}.

The article is organized as follows. In section~2, we present a
brief discussion of some basic properties of pseudo-unitary
operators and their relevance to symplectic transformations. In
section~3, we explore block-diagonalizable pseudo-unitary
operators with finite-dimensional diagonal blocks. In section~4,
we use the results of sections 2 and 3 to study pseudo-unitary
matrices. In section~5, we offer a thorough discussion of the
$2\times 2$ pseudo-unitary matrices. In section~6, we study an
application of our general results for a quantum system with a
pseudo-unitary dynamical group and elaborate on the relation
between the choice of the dynamical group and the issue of second
quantization. Finally, in section~7, we provide a survey of our
main results and present our concluding remarks.

\section{Pseudo-Hermitian and Pseudo-Unitary operators}

By definition \cite{p1}, a linear operator $H:{\cal H}\to{\cal H}$
acting in a Hilbert space ${\cal H}$ is said to be
pseudo-Hermitian if there exists a linear, invertible, Hermitian
operator $\eta:{\cal H}\to{\cal H}$ such that
    \be
    H^\dagger=\eta H \eta^{-1}.
    \label{ph}
    \end{equation}
For a given pseudo-Hermitian operator $H$, the operator $\eta$
satisfying (\ref{ph}) is not unique \cite{p7,cqg}. Each choice of
$\eta$ determines a possibly indefinite inner product (a
pseudo-inner product) on ${\cal H}$, namely
    \be
    \bbr\psi,\phi\kkt_\eta:=\br\psi|\eta\phi\kt,
    \label{inn}
    \end{equation}
where $ \psi,\phi\in{\cal H}$, and $\br~~|~~\kt$ is the original
inner product of ${\cal H}$. Conversely, every pseudo-inner
product on ${\cal H}$ has the form~(\ref{inn}). As a result,
$\eta$ is sometimes called a metric operator.

If we make a particular choice for $\eta$, we say that $H$ is
$\eta$-pseudo-Hermitian. In this case, it is Hermitian with
respect to the inner product $\bbr~~,~~\kkt_\eta$. Therefore, the
study of $\eta$-pseudo-Hermitian operators is equivalent to the
study of Hermitian operators in a vector space with an indefinite
metric \cite{bognar}. The application of the latter in quantum
physics dates back to the 1940s \cite{pauli}. See also
\cite{mielnik,cjp}. As emphasized in \cite{cjp}, there is an
important distinction between the concept of pseudo-Hermiticity,
where one does not fix the inner product and has the freedom of
choosing it, and the well-studied notion of
$\eta$-pseudo-Hermiticity.

We can express the defining condition~(\ref{ph}) in the form
$H^\sharp=H$ where $H^\sharp:= \eta^{-1}H^\dagger\eta$ is the
$\eta$-pseudo-adjoint of $H$. Using the latter one can also define
the notion of an $\eta$-pseudo-unitary operator $U:{\cal
H}\to{\cal H}$ by requiring that $U$ satisfies $U^\sharp=U^{-1}$.
    \begin{itemize}
    \item[] {\bf Definition:} A linear invertible operator
    $U:{\cal H}\to{\cal H}$ is said to be pseudo-unitary if there
    exists a linear, invertible, Hermitian operator
    $\eta:{\cal H}\to{\cal H}$ such that $U$ is
    $\eta$-pseudo-unitary, i.e.,
        \be
        U^\dagger=\eta U^{-1} \eta^{-1}.
        \label{pu}
        \end{equation}
    \end{itemize}
Similarly to the case of pseudo-Hermitian operators, $\eta$ is not
unique. If we make a choice for $\eta$, we say that $U$ is
$\eta$-pseudo-unitary. In this case it is not difficult to show
that $U$ leaves the pseudo-inner product $\bbr~~,~~\kkt_\eta$
invariant. This is easily seen by writing~(\ref{pu}) in the form
    \be
    U^\dagger\eta U=\eta,
    \label{pu2}
    \end{equation}
and using (\ref{inn}) and (\ref{pu2}) to check that
    \be
    \bbr U\psi,U\phi\kkt_\eta=\bbr\psi,\phi\kkt_\eta,~~~~~~~~~~\forall\psi,\phi\in{\cal H}.
    \label{inv}
    \end{equation}

Given an $\eta$-pseudo-Hermitian operator $H$ one can construct a
one-parameter family of $\eta$-pseudo-unitary operators, namely
$U(t)=e^{-it H}$ with $t\in\R$.
    \begin{itemize}
    \item[] {\bf Proposition~1:} Let $\epsilon\in\R^+$,
    $t\in(-\epsilon,\epsilon)$, $H:{\cal H}\to{\cal H}$ be a
    $t$-independent linear operator acting in a Hilbert space ${\cal H}$,
    $U(t):=e^{-it H}$, and $\eta:{\cal H}\to{\cal H}$ be a
    $t$-independent Hermitian, invertible, linear operator.
    Then $H$ is $\eta$-pseudo-Hermitian if and only if $U(t)$ is
    $\eta$-pseudo-unitary for all $t\in(-\epsilon,\epsilon)$.
    \item[] {\bf Proof:} Suppose that $H$ is
    $\eta$-pseudo-Hermitian, then a direct application of
    Eqs.~(\ref{ph}), $U(t)^\dagger=e^{itH^\dagger}$, and
    $U(t)^{-1}=e^{itH}$ shows that $U(t)$ satisfies (\ref{pu}), i.e.,
    it is $\eta$-pseudo-unitary. Conversely, let U(t) be
    $\eta$-pseudo-unitary for all $t\in(-\epsilon,\epsilon)$. Then
    substituting $U(t)$ for $U$ in Eq.~(\ref{pu}), taking the
    derivative of both sides with respect to $t$, and setting $t=0$
    in the resulting expression, we find that $H$ satisfies
    (\ref{ph}), i.e., it is $\eta$-pseudo-Hermitian.~~~$\square$
    \end{itemize}
Because $U(t)$ may be identified with the evolution operator for a
quantum system having $H$ as its Hamiltonian, a quantum system
with a time-independent Hamiltonian has a pseudo-unitary evolution
if and only if the Hamiltonian is pseudo-Hermitian, \cite{cqg}.

The one-parameter family $U(t)$ clearly forms an Abelian Lie group
under composition. This is indeed a subgroup of the group ${\cal
U}_\eta({\cal H})$ of all $\eta$-pseudo-unitary operators. The
latter forms a group because for any pair $U_1,U_2:{\cal
H}\to{\cal H}$ of $\eta$-pseudo-unitary operators,
    \[ (U_1^{-1}U_2)^\dagger=U_2^\dagger (U_1^\dagger)^{-1}=
    \eta U_2^{-1}\eta^{-1}(\eta U_1^{-1}\eta^{-1})^{-1}=
    \eta U_2^{-1}\eta^{-1}\eta U_1\eta^{-1}=
    \eta (U_1^{-1}U_2)^{-1}\eta^{-1}.\]
Therefore ${\cal U}_\eta({\cal H})$ is a subgroup of the group
$GL({\cal H})$ of all invertible linear transformations acting in
${\cal H}$. In \cite{aj}, the authors considered this group for
the case ${\cal H}=\C^n$. They call it `the pseudo-unitary group'.
This terminology is rather misleading as it does not reflect the
important fact that a particular choice for $\eta$ has been made.
In fact, it is not true that the product of any two pseudo-unitary
operators $V_1$ and $V_2$ is pseudo-unitary. This is because they
may belong to ${\cal U}_\eta({\cal H})$ with different $\eta$.
This observation calls for a more careful study of the structure
of the set ${\cal U}({\cal H}):=\cup_{\eta}\,{\cal U}_\eta({\cal
H})$ of all pseudo-unitary operators acting in ${\cal H}$.

In the remainder of this section we discuss two simple properties of
pseudo-unitary operators that will be of future use.
\begin{itemize}
    \item[] {\bf Proposition~2:} Let $\eta_1$ be a Hermitian, invertible,
    linear operator acting in a Hilbert space ${\cal H}$,
    $A:{\cal H}\to{\cal H}$, $U_1:{\cal H}\to{\cal H}$ be invertible
    linear operators, $U_2:=A^{-1}U_1A$ and
    $\eta_2:=A^\dagger\eta_1 A$. Then $U_1$ is
    $\eta_1$-pseudo-unitary if and only if $U_2$ is
    $\eta_2$-pseudo-Hermitian.
    \item[] {\bf Proof:} First note that the defining condition
    (\ref{pu}) may be written in the form
    $U\eta^{-1}U^\dagger\eta=I$, where $I$ is the identity operator.
    Then a simple calculation shows that
        $$U_2\eta_2^{-1}U_2^{\dagger}\eta_2=A^{-1}U_1A
        A^{-1}\eta_1^{-1}A^{-1\dagger}A^\dagger U_1 A^{-1\dagger}
        A^\dagger\eta_1A=A^{-1}(U_1\eta_1^{-1}U_1^{\dagger}\eta_1)A.$$
    Therefore, $U_1\eta_1^{-1}U_1^{\dagger}\eta_1=I$ if and only
    if $U_2\eta_2^{-1}U_2^{\dagger}\eta_2=I$.~~~$\square$
    \item[] {\bf Proposition~3:} Let $U_1:{\cal H}\to{\cal H}$ be
    a pseudo-unitary operator acting in a Hilbert space ${\cal H}$ and
    $u$ be an eigenvalue of $U$. Then $1/u^*$ is also an eigenvalue
    of $U$. In other words, eigenvalues of $U$ are either unimodular
    ($|u|=1$) or they come in inverse-complex-conjugate pairs $(u,1/u^*)$.
    \item[] {\bf Proof:} Let $|u\kt$ be an eigenvector of $U$ with
    eigenvalue $u$, i.e., $U|u\kt=u|u\kt$. Acting both sides of (\ref{pu2})
    on $u^{-1}|u\kt$, we find $U^\dagger\eta|u\kt=u^{-1}\eta|u\kt$. Because
    $\eta$ is invertible, $\eta|u\kt\neq 0$. This in turn means that $u^{-1}$ is
    an eigenvalue of $U^\dagger$. But the eigenvalues of $U^\dagger$ are
    complex-conjugates of those of $U$. Therefore, $u^{-1*}=1/u^*$ is an
    eigenvalue of $U$. If $u=1/u^*$, $u$ is unimodular; otherwise $(u,1/u^*)$
    is a pair of distinct inverse-complex-conjugate eigenvalues.~~~$\square$
    \end{itemize}

As a straightforward application of Proposition~3, consider the
case that ${\cal H}=\C^{2m}$, for some $m\in\Z^+$, and endow
$\C^{2m}$ with the metric operator
    \be
    \eta_{_J}:=iJ,
    \label{eta=j}
    \ee
where $J:\C^{2m}\to\C^{2m}$ has the following matrix
representation in the standard orthonormal basis of $\C^{2m}$.
    \be
    J=\left(\begin{array}{cc}
    0_m & -1_m\\
    1_m & 0_m\end{array}\right).
    \label{eta-symp}
    \ee
Here $0_m$ and $1_m$ are respectively the $m\times m$ zero and
identity matrices respectively. According to (\ref{eta=j}) and
(\ref{eta-symp}), the operator $\eta_{_J}$ has a Hermitian matrix
representation in an orthonormal basis, and $\eta_{_J}^2=1$. Hence
$\eta_{_J}$ is indeed a Hermitian invertible (metric) operator
acting in $\C^{2m}$

Next, observe that the operator $J$ restricted to $\R^{2m}$ yields
the usual symplectic form \cite{arnold} on $\R^{2m}$. The
associated symplectic transformations coincide with real $2m\times
2m$ matrices $S$ satisfying \cite{arnold}
    \be
    S^t\, J~ S=J,
    \label{symp-trans}
    \ee
where $S^t$ stands for the transpose of $S$. We can view the
symplectic transformations $S$ as linear operators acting in
$\C^{2p}$. Then the condition that they admit real matrix
representations (in the standard basis) takes the form
    \be
    T~S~T=S,
    \label{tst}
    \ee
where $T$ is the (time-reversal) operator defined by $\forall \vec
z\in\C^{2p}$, $T\vec z=\vec z^*$. Making use of (\ref{eta=j}) and
the fact that $T^{-1}=T$ and $S^\dagger=S^t$, we can respectively
express the defining relations (\ref{symp-trans}) and (\ref{tst})
of the symplectic transformations $S$ as
    \bea
    S^\dagger\eta_{_J} S &=&\eta_{_J},
    \label{ses=e} \\
    \left[S,T\right] &=&0.
    \label{st=0}
    \eea
Because $T$ is an antilinear Hermitian invertible operator,
according to Theorem~2 of Ref.~\cite{p3}, Eq.~(\ref{st=0}) implies
that $S$ is a pseudo-Hermitian operator. Furthermore,
Eq.~(\ref{ses=e}) means that $S$ is in addition a pseudo-unitary
operator.

In view of Proposition~3 and the spectral characterization theorem
for pseudo-Hermitian operators \cite[Theorem~2]{p1}, the fact that
symplectic transformations are both pseudo-Hermitian and
pseudo-unitary leads to the following well-known spectral theorem
for symplectic transformations~\cite{arnold}.
    \begin{itemize}
    \item[] {\bf Theorem~1:} Let $\lambda$ be an eigenvalue of a
    symplectic transformation~$S$, then so are $\lambda^*$,
    $1/\lambda$, and $1/\lambda^*$.
    \item[] {\bf Proof:} Because $S$ is pseudo-unitary
    $1/\lambda^*$ is an eigenvalue. Because it is pseudo-Hermitian
    $\lambda^*$ and $(1/\lambda^*)^*=1/\lambda$ are
    eigenvalues.~~~$\square$
    \end{itemize}

\section{Block-Diagonalizable Pseudo-Unitary Operators with
Finite-Dimensional Diagonal Blocks}

Consider an operator $U:{\cal H}\to{\cal H}$ acting in a Hilbert
space ${\cal H}$ and having a discrete spectrum. Then $U$ is said
to be block-diagonalizable with finite-dimensional diagonal blocks
\cite{p6} if it can be expressed in the form
    \be
    U=\sum_n \sum_{a=1}^{d_n}\left(u_n\sum_{i=1}^{p_{n,a}}
    |\psi_n,a,i\kt\br\phi_n,a,i|+\sum_{i=1}^{p_{n,a}-1}
    |\psi_n,a,i\kt\br\phi_n,a,i+1|\right),
    \label{block-u}
    \end{equation}
where $n$ is the spectral label, $u_n$ are the eigenvalues of $U$,
$d_n$ is the geometric multiplicity of $u_n$,
$a\in\{1,2,\cdots,d_n\}$ is a degeneracy label, $p_{n,a}$ is the
dimension of the Jordan block associated with the labels $n$ and
$a$ (These are called the Jordan dimensions \cite{p6}), and
$\{|\psi_n,a,i\kt,|\phi_n,a,i\kt\}$ is a complete biorthonormal
system satisfying
    \be
    \br\psi_n,a,i|\phi_m,b,j\kt=\delta_{mn}\delta_{ab}\delta_{ij},
    ~~~~~~~
    \sum_n\sum_{a=1}^{d_n}\sum_{i=1}^{p_{n,a}}
    |\psi_n,a,i\kt\br\phi_m,b,j|=1.
    \label{bior}
    \end{equation}
In view of (\ref{block-u}) and (\ref{bior}),
    \be
    U|\psi_n,a,1\kt=u_n|\psi_n,a,1\kt,~~~~~~~
    U^\dagger|\phi_n,a,p_{n,a}\kt=u_n^*|\phi_n,a,p_{n,a}\kt,
    \label{eg-va-u}
    \end{equation}
i.e., $|\psi_n,a,1\kt$ are the eigenvectors of $U$ and
$|\phi_n,a,p_{n,a}\kt$ are the eigenvectors of $U^\dagger$.
Clearly, the eigenvalues of $U^\dagger$ are complex-conjugates of
those of $U$, and if $U$ is invertible the eigenvalues $u_n$ do
not vanish.

\begin{itemize}
\item[] {\bf Lemma~1:} Let $U:{\cal H}\to{\cal H}$ be an
invertible operator acting in a Hilbert space ${\cal H}$ and
$z\in\C-\{0\}$. Then for all $\ell\in\Z^+$,
    \be
    {\rm kernel}[(U^{-1}-z^{-1})^\ell]={\rm kernel}[(U-z)^\ell].
    \label{L1}
    \end{equation}
\item[] {\bf Proof:} This identity follows by induction over
$\ell$. For $\ell=1$, we have
    \bea
    |\xi\kt\in{\rm kernel}[U^{-1}-z^{-1}]
    &\dto &(U^{-1}-z^{-1})|\xi\kt=0\nn\\
    &\dto & z U(U^{-1}-z^{-1})|\xi\kt=0\nn\\
    &\dto & (z-U)|\xi\kt=0\nn\\
    &\dto &|\xi\kt\in{\rm kernel}[U-z],
    \label{L2}
    \eea
where we have used the fact that $z U$ is an invertible operator.
(\ref{L2}) shows that (\ref{L1}) holds for $\ell=1$. Now, suppose
(\ref{L1}) holds for some $\ell=k\in\Z^+$. Then
    \bea
    |\xi\kt\in {\rm kernel}[(U^{-1}-z^{-1})^{k+1}]
    &\dto & (U^{-1}-z^{-1})^{k}(U^{-1}-z^{-1})|\xi\kt=0\nn\\
    &\dto & (U^{-1}-z^{-1})|\xi\kt
        \in{\rm kernel}[(U^{-1}-z^{-1})^{k}]\nn\\
    &\dto & (U^{-1}-z^{-1})|\xi\kt\in{\rm kernel}[(U-z)^k]\nn\\
    &\dto & (U-z)^k(U^{-1}-z^{-1})|\xi\kt=0\nn\\
    &\dto & z U(U-z)^k(U^{-1}-z^{-1})|\xi\kt=0\nn\\
    &\dto & (U-z)^k(z-U)|\xi\kt=0\nn\\
    &\dto & |\xi\kt\in {\rm kernel}[(U-z)^{k+1}].\nn
    \eea
Therefore, (\ref{L1}) holds for $\ell=k+1$; by induction, it holds for
all $\ell\in\Z^+$.~~~$\square$
    \item[] {\bf Theorem~2:} Let $U:{\cal H}\to{\cal H}$ be an
    operator acting in a Hilbert space ${\cal H}$ and having a
    discrete spectrum. Suppose that $U$ is block-diagonalizable
    with finite-dimensional diagonal blocks so that
    (\ref{block-u}) holds. Then $U$ is pseudo-unitary if and only
    if the eigenvalues $u_n$ of $U$ are either unimodular (i.e.,
    $|u_n|=1$) or they come in inverse-complex-conjugate pairs
    $(u_n,1/u_n^*)$ and that the geometric multiplicity and the
    Jordan dimensions for the inverse-complex-conjugate eigenvalues
    coincide.
    \item[] {\bf Proof:} Suppose that $U$ is pseudo-unitary. Then,
    according to Proposition~3  the eigenvalues of $U$ are either
    unimodular or they come in inverse-complex-conjugate pairs.
    Suppose that $u_n$ and $1/u_n^*$ form a pair of distinct
    inverse-complex-conjugate eigenvalues. In order to show
    that they have the same geometric multiplicity and Jordan
    dimensions we prove that for all $\ell\in\Z^+$, ${\rm
    kernel}(U-u_n)^\ell$ and ${\rm kernel}(U-1/u_n^*)^\ell$ have
    the same (finite) dimension. To see this, first note that
    $U$ and $U^\dagger$ have the same Jordan block structure; in
    view of (\ref{block-u}), for all $\ell\in\Z^+$, ${\rm
    kernel}(U-u_n)^\ell$ and ${\rm kernel}(U^\dagger-u_n^*)^\ell$
    have the same (finite) dimension. Hence they are isomorphic as
    vector spaces. Next, we use the fact that $\eta$ is an
    invertible operator to establish the isomorphism between
    ${\rm kernel}(U^\dagger-u_n^*)^\ell$ and
        \bea
        {\rm kernel}[\eta^{-1}(U^\dagger-u_n^*)^\ell\eta]
        &=&{\rm kernel}[(\eta^{-1}U^\dagger\eta-u_n^*)^\ell]\nn\\
        &=&{\rm kernel}[(U^{-1}-u_n^*)^\ell]\nn\\
        &=&{\rm kernel}[(U-1/u_n^*)^\ell].\nn
        \eea
    Here we have made use of the defining relation~(\ref{pu}) and
    the identity (\ref{L1}) of Lemma~1. This completes the proof
    that for all $\ell\in\Z^+$, ${\rm kernel}(U-u_n)^\ell$ is
    isomorphic to ${\rm kernel}(U-1/u_n^*)^\ell$. Therefore, they
    have the same (finite) dimension.

    Next, suppose that $U$ has unimodular and/or
    inverse-complex-conjugate pairs of eigenvalues with identical
    geometric multiplicity and Jordan dimensions. Then $U$ may be
    expressed as
    \bea
    U&=&\sum_{\nu_0}\sum_{a=1}^{d_{\nu_0}}
        \left( u_{\nu_0}\sum_{i=1}^{p_{\nu_0,a}} |\psi_{\nu_0},a,i\kt\br\phi_{\nu_0},a,i|+
        \sum_{i=1}^{p_{\nu_0,a}-1} |\psi_{\nu_0},a,i\kt\br\phi_{\nu_0},a,i+1|\right)+\nn\\
       && \sum_{\nu}\sum_{a=1}^{d_{\nu}}\left[
        \sum_{i=1}^{p_{\nu,a}}\left( u_{\nu}|\psi_{\nu+},a,i\kt\br\phi_{\nu+},a,i|+
        \frac{1}{u_\nu^*} |\psi_{\nu-},a,i\kt\br\phi_{\nu-},a,i|\right)+\right.\nn\\
       && \left.\sum_{i=1}^{p_{\nu,a}-1} \left(
         |\psi_{\nu+},a,i\kt\br\phi_{\nu+},a,i+1| +|\psi_{\nu-},a,i\kt\br\phi_{\nu-},a,i+1|\right)
        \right],
    \label{u1}
    \eea
where we have set $n=\nu_0,\nu+$, or $\nu-$ depending
on whether $|u_n|=1$, $|u_n|>1$, or $|u_n|<1$ respectively,
and used $\nu$ to denote the common value of $\nu+$ and
$\nu-$. In order to show that $U$, as given by (\ref{u1}),
is pseudo-unitary we construct a Hermitian, invertible, linear
operator $\eta$ satisfying (\ref{pu}) or equivalently (\ref{pu2}).
Consider the ansatz
    \bea
    \eta &=&\sum_{\nu_0}\sum_{a=1}^{d_{\nu_0}}\sum_{i,j=1}^{p_{\nu_0,a}}
            z_{\nu_0,a,i,j}  |\phi_{\nu_0},a,i\kt\br\phi_{\nu_0},a,j|+\nn\\
        && \sum_{\nu}\sum_{a=1}^{d_{\nu}}\sum_{i,j=1}^{p_{\nu,a}}
            \left( \zeta_{\nu,a,i,j}  |\phi_{\nu-},a,j\kt\br\phi_{\nu+},a,i|+
            \zeta_{\nu,a,i,j}^*  |\phi_{\nu+},a,i\kt\br\phi_{\nu-},a,j|\right),
    \label{eta1}
    \eea
where $z_{\nu_0,a,i,j}$ and $\zeta_{\nu,a,i,j}$ are complex coefficients and
    \be
    z_{\nu_0,a,i,j}^*=z_{\nu_0,a,j,i}.
    \label{condi1}
    \end{equation}
The latter relation ensures that $\eta$ is Hermitian. Now, impose
the condition (\ref{pu2}). Substituting (\ref{u1}) and
(\ref{eta1}) in (\ref{pu2}) and using the biorthonormality and
completeness relations (\ref{bior}), we find after a quite lengthy
calculation that $z_{\nu_0,a,i,j}$ and $\zeta_{\nu,a,i,j}$ are
solutions of the following equations for
$u=u_{\nu_0},p=p_{\nu_0,a}$ and $u=u_{\nu},p=p_{\nu,a}$,
respectively.
    \bea
    &&x_{1,i}=x_{i,1}=0,~~~~~~~~~~~~~~~~~~~~~~~~~~~\forall i\in\{1,2,\cdots,p-1\},
    \label{x1}\\
    &&u\,x_{i-1,j}+u^{-1}x_{i,j-1}+x_{i-1,j-1}=0,~~~~~~~~~~~\forall i,j\in\{2,\cdots,p\}.
    \label{x2}
    \eea
It turns out that these equations have the following exact solution:
    \be
    x_{i,j}=\left\{
    \begin{array}{c}
    0 ~~~~~~~~~~~~~~~~~~ {\rm for} ~~~~~~~~~~~~~~ i+j \leq p\\
    \\
    \sum_{k=1}^{i+j-p}\mbox{\scriptsize $\left(\begin{array}{c}
        i-k-1\\
        p-j-1\end{array}\right)$}\, (-1)^{i-k} u^{p+i-j-k} x_{k,p}~~~~ {\rm for} ~~~~ j<p<i+j,
    \end{array}\right.
    \label{x=}
    \end{equation}
where for all $r,s\in\Z^+$ with $r\leq s$
    \[\mbox{\scriptsize $ \left(\begin{array}{c}
        s\\
        r\end{array}\right)$}:=\frac{s !}{r!(s-r)!},\]
and $x_{k,p}$ with $k\in\{1,2,\cdots,p\}$ are arbitrary complex numbers. We have obtained the solution~(\ref{x=}) by a tedious inspection scheme and checked its validity by direct substitution in (\ref{x2}); it clearly satisfies (\ref{x1}). It is important to note that according to ~(\ref{x=}), $x_{i,j}$ form a $p\times p$ matrix $x$ of the form
    \be
    x=\left(\begin{array}{cccccccc}
    0 & 0 & 0 & \cdots  & 0 & 0 & x_{1,p}\\
    0 & 0 & 0 & \cdots  & 0 & x_{2,p-1} & x_{2,p} \\
    0 & 0 & 0 & \cdots  & x_{3,p-2} & x_{3,p-1} & x_{3,p} \\
    \vdots &\vdots &\vdots &\vdots\; \vdots\; \vdots  & \vdots  & \vdots  & \vdots \\
    0 & 0 & x_{p-2,3} & \cdots & x_{p-2,p-2}& x_{p-2,p-1} & x_{p-2,p} \\
    0 & x_{p-1,2}& x_{p-1,3} & \cdots & x_{p-1,p-2}& x_{p-1,p-1} & x_{p-1,p} \\
    x_{p,1}& x_{p,2}& x_{p,3}& \cdots & x_{p,p-2}& x_{p,p-1}  & x_{p,p}
    \end{array}\right).
    \label{X}
    \end{equation}
In view of (\ref{x=}) all the entries of $x$ are determined in terms of the entries in the last column.
For example, we have
    \be
    x_{i,p-i+1}=(-1)^{i-1} u^{2(i-1)} x_{1,p},~~~~~~~~~~~\forall i\in
    \{1,2,\cdots,p\}.
    \label{x_diag}
    \end{equation}
Moreover note that the determinant of $x$ is up to a sign the
product of the entries (\ref{x_diag}). Therefore, $x$ is an
invertible matrix provided that $x_{1,p}\neq 0$ and $u\neq 0$.
Next, consider the case that $u$ is unimodular and seek for the
solutions (\ref{x=}) that make $x$ Hermitian, i.e., find solutions
for (\ref{x1}) and (\ref{x2}) subject to the condition
    \be
    x_{i,j}^*=x_{j,i}.
    \label{hermit}
    \end{equation}
Imposing this condition on the solution (\ref{x=}) restricts the
choice of the initially free entries, namely $x_{i,p}$. For
example, setting $i=p$ and $j=1$ in (\ref{x=}) or alternatively
setting $i=p$ in (\ref{x_diag}), we find $x_{p,1}=(-1)^{p-1}
u^{2(p-1)} x_{1,p}$. Now, using (\ref{hermit}) which implies
$x_{p,1}=x_{1,p}^*$, we find
    \be
    x_{1,p}=\pm \sqrt{(-1)^{p-1}}\; u^{1-p} \rho,
    \label {x1p}
    \end{equation}
where $\rho=|x_{1,p}|$ is an arbitrary nonnegative real number. A
similar analysis shows that the condition (\ref{hermit}) leads to
similar restrictions on the choices of $x_{i,p}$ with $i>1$. But
these restrictions do not lead to any contradictions, i.e.,
(\ref{hermit}) can always be satisfied. Indeed there are
infinitely many solutions of the form (\ref{x=}) that fulfil
(\ref{hermit}). In particular, if we choose $|u|=1$ and $\rho\neq
0$, the matrix $x$ is an invertible Hermitian matrix. Setting
$u=u_{\nu_0}$, we have a set of solutions $z_{\nu_0,a,i,j}$ of
(\ref{x1}) and (\ref{x2}) that respect the condition
(\ref{condi1}) and that the matrices $z_{\nu_0,a}$ formed out of
$z_{\nu_0,a,i,j}$ are invertible. Similarly, setting $u=u_{\nu}$
we have a set of solutions $\zeta_{\nu,a,i,j}$ of (\ref{x1}) and
(\ref{x2}) such that the matrices $\zeta_{\nu,a}$ formed out of
$\zeta_{\nu,a,i,j}$ are also invertible. The existence of these
solutions is equivalent to the existence of a linear operator
$\eta$ of the form (\ref{eta1}) that satisfies (\ref{pu2}) and is
Hermitian and invertible. The inverse of $\eta$ is given by
    \bea
    \eta^{-1} &=&\sum_{\nu_0}\sum_{a=1}^{d_{\nu_0}}\sum_{i,j=1}^{p_{\nu_0,a}}
            \tilde z_{\nu_0,a,i,j}  |\psi_{\nu_0},a,i\kt\br\psi_{\nu_0},a,j|+\nn\\
        && \sum_{\nu}\sum_{a=1}^{d_{\nu}}\sum_{i,j=1}^{p_{\nu,a}}
            \left( \tilde\zeta_{\nu,a,i,j}  |\psi_{\nu-},a,j\kt\br\psi_{\nu+},a,i|+
            \tilde\zeta_{\nu,a,i,j}^*  |\psi_{\nu+},a,i\kt\br\psi_{\nu-},a,j|\right),
    \label{eta-inv}
    \eea
where $\tilde z_{\nu_0,a,i,j}$ are the entries of the matrix
$z_{\nu_0,a}^{-1}$, and $\tilde\zeta_{\nu,a,i,j}$ are those of
$\zeta_{\nu,a}^{-1\dagger}$. One can check by direct calculation
that $\eta^{-1}\eta=1$. This completes the proof of the
pseudo-unitarity of $U$. ~~~$\square$
    \end{itemize}

\section{Pseudo-Unitary Matrices}

According to Theorem~2, a square matrix $U$ is pseudo-unitary if
its eigenvalues are either unimodular or they come in
inverse-complex-pairs and that geometric multiplicity and the
Jordan dimensions of the latter are identical. A direct
consequence of this observation is the following.
    \begin{itemize}
    \item[] {\bf Proposition~4:} Every pseudo-unitary matrix $U$
    has a unimodular determinant, i.e., $|\det U|=1$.
    \item[] {\bf Proof:} This follows from the fact that in the
    Jordan canonical form of $U$ the nonunimodular entries come is
    inverse-complex-conjugate pairs, $(u_n,1/u_n^*)$. Hence their
    product which yields $\det U$ is unimodular.~~~$\square$
    \end{itemize}
According to this proposition the set ${\cal U}(\C^n)$ of all
$n\times n$ pseudo-unitary matrices is a subset of the group
    \be
    \Sigma L(n,\C):=\{g\in GL(n,\C)|~~|\det g|=1\},
    \label{ps-group}
    \end{equation}
of $n\times n$ matrices with unimodular determinant. We shall call
$\Sigma L(n,\C)$ the {\em pseudo-special groups}. As a subset of
$GL(n,\C)$, $\Sigma L(n,\C)$ is the inverse image of the group
$U(1)$ under the homomorphism $\det: GL(n,\C)\to GL(1,\C)$.
Therefore, $\Sigma L(n,\C)$ is a subgroup of $GL(n,C)$. In fact,
it is not difficult to show that $\Sigma L(n,\C)$ is isomorphic to
the product group $U(1)\times SL(n,\C)$. Note however that not
every element of the pseudo-special groups is pseudo-unitary. For
example let $g$ be a $2\times 2$ diagonal matrix with diagonal
entries $2i$ and $-i/2$. Clearly, $\det g = 1\in U(1)$, so $g\in
\Sigma L(2,\C)$. But, $(2i)^{-1*}=i/2\neq -i/2$. Hence the
eigenvalues $2i$ and $-i/2$ are not inverse-complex-conjugates,
and $g$ is not pseudo-unitary. In general, ${\cal U}(\C^n)$ is a
proper subset of $\Sigma L(n,\C)$.

Next, consider the group ${\cal U}_{\eta}\,(\C^n)$ for a fixed
Hermitian invertible $n\times n$ matrix $\eta$. We recall
Sylvester's law of inertia according to which $\eta$ satisfies
    \be
    \eta=A^\dagger\eta_{p,q} A,
    \label{syl}
    \end{equation}
where $A$ is some invertible $n\times n$ matrix and $\eta_{p,q}$
is a diagonal matrix of the form
    \be
    \eta_{p,q}={\rm diag}(-1,-1,\cdots,-1,1,1,\cdots,1),
    \label{eta-can}
    \end{equation}
which has $p$ negative and $q:=n-p$ positive entries.
    \begin{itemize}
    \item[] {\bf Proposition~5:} Let $\eta$ be an $n\times n$
    Hermitian and invertible matrix. Then the group
    ${\cal U}_\eta(\C^n)$ is isomorphic to the `pseudo-unitary'
    group
        $$ U(p,q):=\left\{ g\in GL(n,\C)|g^\dagger\eta_{p,q}g=
        \eta_{p,q}\right\}={\cal U}_{\eta_{p,q}}(\C^n),$$
    for some $p\in\{0,1,\cdots,n\}$ and $q:=n-p$.%
    \footnote{Note that $U(0,n)=U(n)$.}
    \item[] {\bf Proof:} Setting $U_2=U$, $\eta_2=\eta$,
    and $\eta_1=\eta_{p,q}$ in Proposition~2, we see that
    $U\in{\cal U}_\eta\,(\C^n)$ if and only if
    $U_1:=A U A^{-1}\in U(p,q)$. Hence,
    ${\cal U}_\eta\,(\C^n)=A^{-1}U(p,q)A$. Because the
    conjugation $i_A:GL(n,\C)\to GL(n,\C)$ defined by
    $i_A(g):=A^{-1}gA$ is an automorphism of the group $GL(n,\C)$
    that maps ${\cal U}_\eta\,(\C^n)$ onto $U(p,q)$, the
    subgroups ${\cal U}_\eta\,(\C^n)$ and $U(p,q)$ are
    isomorphic.~~~$\square$
    \end{itemize}

According to Proposition~5, the pseudo-unitary groups ${\cal
U}_\eta(\C^n)$ are isomorphic to and obtained from the classical
groups $U(p,q)$ (or $U(n)$) by conjugation; ${\cal
U}_\eta(\C^n)=A^{-1}U(p,q)A$ for some $A\in GL(n,\C)$. Therefore,
the set ${\cal U}(\C^n)$ may be viewed as the union of the orbits
of the subgroups $U(p,q)$ under conjugation in $GL(n,\C)$.
Obviously these orbits, which according to Proposition~4 lie in
the pseudo-special group $\Sigma L(n,\C)$, are not disjoint. For
example, $e^{iH}\in{\cal U}(\C^n)$ belongs to both ${\cal
U}_{\eta_1}(\C^n)$ and ${\cal U}_{\eta_2}(\C^n)$, if $H$ is both
$\eta_1$- and $\eta_2$-pseudo-Hermitian. The latter holds if and
only if $\eta_2=A^\dagger\eta_1 A$ for some $A\in GL(n,\C)$
commuting with $H$, \cite{p7}.

Another simple consequence of Proposition~5 is the following.
    \begin{itemize}
    \item[] {\bf Corollary:} Let $m\in\Z^+$. Then the group $Sp(2m)$
    of symplectic transformations of $\R^{2m}$ is isomorphic to the
    real subgroup of (a matrix group that is isomorphic to) the
    pseudo-unitary group $U(m,m)$.
    \item[] {\bf Proof:} According to the argument given above
    Theorem~1, $Sp(2m)$ may be identified with the subgroup of
    ${\cal U}_{\eta_{_J}}(\C^{2m})$ consisting of real matrices.
    It is not difficult to show that the spectrum of $\eta_{_J}$
    consists of $-1$ and $1$ each with multiplicity $m$. Hence
    according to Proposition~5, ${\cal U}_{\eta_{_J}}(\C^{2m})$ is
    isomorphic to $U(m,m)$, and $Sp(2m)$ is isomorphic to the real
    subgroup of this group.~~~$\square$
    \end{itemize}
Note also that according to the argument used in the above proof
of Theorem~1 and the spectral characterization theorems for
pseudo-Hermitian and pseudo-unitary operators (i.e., Theorem~1 of
Ref.~\cite{p6} and Theorem~2 above), given an eigenvalue $\lambda$
of a symplectic transformation $S\in Sp(2m)$, the eigenvalues
$\lambda^*,1/\lambda,$ and $1/\lambda^*$ have the same geometric
multiplicity and Jordan dimensions as $\lambda$. This in
particular proves the well-known fact that $S$ has a unit
determinant. In particular, $Sp(2m)$ may be identified with the
real subgroup of (a matrix group that is isomorphic to) $SU(m,m)$.

Next, we state and prove the following lemma.
    \begin{itemize}
    \item[] {\bf Lemma~2:} Let $p\in\Z^+$, $E\in\C$, and $h$ be a
    $p\times p$ matrix of the Jordan form
        \be
        h=E 1_p+a_p,
        \label{L2-1}
        \end{equation}
    where $1_p$ is the $p\times p$ identity matrix and $a_p$ is
    the  $p\times p$ matrix\footnote{$a_p$ provides an
    irreducible representation of the annihilation operator for a
    parafermion of order $p-1$, \cite{ok}.}
        \be
        a_p:= \left(\begin{array}{ccccccc}
        0 & 1 & 0 & \cdots & 0 & 0 & 0\\
        0 & 0 & 1 & \cdots & 0 & 0 & 0\\
        \vdots & \vdots & \ddots
        &\ddots~\ddots&\vdots&\vdots&\vdots\\
        0 & 0 & 0 & \cdots & 0 & 1 & 0 \\
        0 & 0 & 0 & \cdots & 0 & 0 & 1 \\
        0 & 0 & 0 & \cdots & 0 & 0 & 0
        \end{array}\right).
        \label{a=}
        \end{equation}
    Then $e^{ih}$ has the following canonical Jordan form,
        \be
        e^{iE}1_p+a_p.
        \label{L2-2}
        \end{equation}
    Equivalently, $e^{iE}$ is the unique eigenvalue of $e^{ih}$
    with geometric multiplicity 1 and algebraic multiplicity $p$.
    \item[]{\bf Proof:} Using the fact that $a_p^p=0$, we can
    easily compute
        \[ e^{ih}=e^{iE}\sum_{\ell=0}^{p-1}
        \frac{i^\ell a^\ell}{\ell !}.\]
    This is an upper triangular matrix with a single eigenvalue
    (namely $e^{iE}$) and a single (linearly independent)
    eigenvector. Therefore its geometric multiplicity is 1 and its
    algebraic multiplicity is $p$.~~~$\square$
    \item[] {\bf Theorem~3:} Every pseudo-unitary matrix $U$ may
    be expressed as $e^{iH}$ for some pseudo-Hermitian matrix
    $H$.
    \item[] {\bf Proof:} Let $U$ be an $n\times n$ pseudo-unitary
    matrix. Clearly, $U\in GL(n,\C)$. Now, because the exponential
    map for the group $GL(n,\C)$ is onto \cite{dk}, there is a
    square matrix $H$ such that $U=e^{iH}$. We can perform a
    similarity transformation $H\to \tilde H:=A^{-1}H A$ that maps
    $H$ into its Jordan canonical form $\tilde H$. We then have
        \be
        U=A e^{i\tilde H}A^{-1}.
        \label{T2-0}
        \end{equation}
    In view of Proposition~2 and Lemma~2, $e^{i\tilde H}$ is
    pseudo-unitary, and its eigenvalues are of the form $e^{i E_n}$
    where $E_n$ are the eigenvalues of $\tilde H$.
    Moreover the geometric multiplicity
    and the Jordan dimensions of (the canonical Jordan form of)
    $e^{i\tilde H}$ coincide with those of $\tilde H$. Now, because
    $e^{i\tilde H}$ is pseudo-unitary, Theorem~2 implies that the
    eigenvalues $e^{iE_n}$ of $e^{i\tilde H}$ are either unimodular
    or they come in inverse-complex-conjugate pairs with identical
    geometric multiplicity and Jordan dimensions. First we
    consider the unimodular eigenvalues which we denote by
    $e^{iE_{\nu_0}}$. Because $1=|e^{iE_{\nu_0}}|^2=
    e^{iE_{\nu_0}}e^{-iE_{\nu_0}^*}$, we have $E_{\nu_0}-E_{\nu_0}^*
    =2\pi k_{\nu_0}$ for some $k_{\nu_0}\in\Z$. But the left-hand
    side  of this equation is imaginary while its right-hand side is
    real. This implies $k_{\nu_0}=0$. Hence $E_{\nu_0}$ is real.
    Next, consider the eigenvalues $e^{iE_{\nu}}$ that are not
    unimodular. These
    are paired with their inverse-complex-conjugate, namely
    $e^{iE_{\nu}^*}$. $e^{iE_{\nu}}$ and $e^{iE_{\nu}^*}$ have
    the same geometric multiplicity $d_\nu$ and Jordan dimensions
    $p_{\nu,a}$. Because $e^{iE_{\nu}^*}$ is an eigenvalue of
    $e^{i\tilde H}$, according to Lemma~2 there is an eigenvalue
    $E_\nu'$ of $\tilde H$ such that
        \be
        e^{iE_{\nu}^*}=e^{iE_{\nu}'},
        \label{T2-1}
        \end{equation}
    and that $E_\nu'$ has the same geometric multiplicity and Jordan
    dimensions as $e^{iE_{\nu}^*}$. Hence the geometric multiplicity
    and Jordan dimensions of $E_\nu'$ are respectively $d_\nu$ and
    $p_{\nu,a}$. Furthermore, Eq.~(\ref{T2-1}) implies
    $E_\nu'=E_\nu^*+2\pi k_\nu$ for some $k_\nu\in\Z$. Now, let
    $E_{\nu+}$ and $E_{\nu-}$
    respectively denote the eigenvalues of $\tilde H$ with positive
    and negative imaginary part. In view of the preceding
    argument, for each $E_{\nu+}$ there is an eigenvalue
    $E_{\nu-}=E_{\nu+}^*+2\pi k_{\nu+}$. Furthermore all the
    eigenvalues with negative imaginary part may be obtained
    from the eigenvalues with positive imaginary part in this way.
    Now, let $\tilde H'$ be the matrix obtained from $\tilde H$ by
    replacing the eigenvalues $E_{\nu-}$ with
    $E_{\nu-}':=E_{\nu-}-2\pi k_{\nu+}=E_{\nu+}^*$. Then,
    by construction, $\tilde H'$ has real and/or complex-conjugate
    pairs of eigenvalues, the latter having identical geometric
    multiplicity and Jordan dimensions. In light of Theorem~1
    of Ref.~\cite{p6}, this implies that $\tilde H'$ is
    pseudo-Hermitian. One can also check that
        \be
        e^{i\tilde H'}=e^{i\tilde H}.
        \label{T2-2}
        \end{equation}
    Next, let
        \be
        H':=A\tilde H' A^{-1}.
        \label{T2-3}
        \end{equation}
    Clearly, $\tilde H'$ is the Jordan canonical form of $H'$. In
    particular, $H'$ is also pseudo-Hermitian. Combining
    Eqs.~(\ref{T2-0}), (\ref{T2-2}), and (\ref{T2-3}), we finally
    have
        \[U=A e^{i\tilde H}A^{-1}=A e^{i\tilde H'}A^{-1}=
        e^{iA\tilde H'A^{-1}}=e^{iH'}.\]
    This completes the proof of the fact that $U$ is the exponential
    of $i$ times a pseudo-Hermitian matrix.~~~$\square$
    \item[] {\bf Corollary~1:} A square matrix $U$ is pseudo-unitary
    if and only if $-i\ln U$ is pseudo-Hermitian, i.e., $U=e^{iH}$
    for a pseudo-Hermitian matrix $H$.
    \item[] {\bf Proof:} If $U$ is pseudo-unitary, then according
    to Theorem~3 it is of the form $e^{iH}$ for some
    pseudo-Hermitian matrix. If $U=e^{iH}$ for a pseudo-Hermitian
    matrix $H$, then setting $\epsilon=2$, $t=-1$ in Proposition~1
    we find that $U=U(-1)$ is pseudo-unitary.~~~$\square$
        \end{itemize}

Corollary~1 is rather surprising, for it is well-known that the
exponential map is not onto for pseudo-unitary groups such as
$U(1,1)$, \cite{dk}. This does not however contradict the
statement of Corollary~1, because when one speaks of a
pseudo-unitary group one fixes the operator $\eta$. What has been
done in the proof of Theorem~3 is to show that for a given
pseudo-unitary operator $U$ there is an $\eta$ such that $U$ is
$\eta$-pseudo-unitary and $H:=-\ln U$ is $\eta$-pseudo-Hermitian.
This is not equivalent to the erroneous statement that given an
$\eta$, $-i\ln U$ is $\eta$-pseudo-Hermitian for every
$\eta$-pseudo-unitary matrix $U$. The exponential map for the
pseudo-unitary group ${\cal U}_\eta(\C^n)$ is generally not onto,
but the exponential map for the set of all pseudo-unitary matrices
is onto. This is another demonstration of the importance of the
difference between the notions of $\eta$-pseudo-Hermiticity
(respectively $\eta$-pseudo-unitarity) and pseudo-Hermiticity
(respectively pseudo-unitarity) \cite{cjp}.

\section{$2\times 2$ Pseudo-Unitary Matrices}

In this section we shall study the case $n=2$ in more detail. The
following corollary of Theorem~3 yields the general form of
$2\times 2$ pseudo-unitary matrices.
\begin{itemize}
\item[] {\bf Corollary~2:} A $2\times 2$ matrix $U$ is
pseudo-Hermitian if and only if $U=A^{-1}D\,A$ where $A$ is an
invertible $2\times 2$ matrix and $D$ is a matrix assuming one of
the following three forms:
    \bea
    D_1&=&\left(\begin{array}{cc}
    e^{i\theta}& 0\\
    0 & e^{i(\varphi-\theta)}\end{array}\right),~~~~~~~~~\theta,
    \varphi\in\R,
    \label{D1}\\
    D_2&=&\left(\begin{array}{cc}
    r e^{i\theta}& 0\\
    0 & e^{i\theta}/r\end{array}\right),~~~~~~~~~r\in\R^+,
    ~\theta\in\R,
    \label{D2}\\
    D_3&=&\left(\begin{array}{cc}
    e^{i\theta}& 1\\
    0 & e^{i\theta}\end{array}\right),~~~~~~~~~~~\theta\in\R.
    \label{D3}
    \eea
\item[] {\bf Proof:} Block diagonalizing $U$ we find a matrix $D$
which is either diagonal or has the form
    \be
    D=\left(\begin{array}{cc}
    u&1\\
    0&u\end{array}\right),
    \label{d=}
    \end{equation}
where $u\in\C$. According to Proposition~2, $D$ is also
pseudo-unitary. This together with Theorem~3 imply that
    \begin{itemize}
    \item if $D$ is diagonal, its eigenvalues are either both
    unimodular, i.e., $D$ is of the form (\ref{D1}), or they are
    inverse-complex-conjugate, i.e., $D$ is of the form (\ref{D2});
    \item If $D$ has the form (\ref{d=}), then it has a single
    eigenvalue $u$ which is necessarily unimodular. That is $D$ is
    of the form (\ref{D3}).~~~$\square$
    \end{itemize}
\end{itemize}
In order to demonstrate the utility of Theorem~3, here we include
a direct proof of Corollary~2. This proof involves the calculation
of the matrices $\eta$ whose general form is given in the proof of
Theorem~2.
    \begin{itemize}
\item[] {\bf A Direct Proof of Corollary~2:} First consider the
case that $U$ is diagonalizable, then the canonical Jordan form
$D=AUA^{-1}$ of $U$ is diagonal. Clearly $\det D=\det U$ and
according to Proposition~4 $\det U\in U(1)$. Hence $|\det D|=1$.
This implies that $D$ must have the form
        \be
        D=\left(\begin{array}{cc}
        \zeta & 0\\
        0 & e^{i\varphi}/\zeta \end{array}\right),
        \label{d}
        \end{equation}
where $\zeta:=r\,e^{i\theta}\in\C-\{0\}$ and $e^{i\varphi}\in
U(1)$, i.e., $r\in\R^+$ and $\theta,\varphi\in\R$. Next, note that
in view of Proposition~2, $U$ is $\eta$-pseudo-unitary if and only
if $D$ is $A^{-1\dagger}\eta A^{-1}$-pseudo-unitary. This reduces
the problem to finding the necessary and sufficient conditions on
$\zeta$ (alternatively $r,\theta$) and $\varphi$ that make $D$
pseudo-unitary. Using the general form
    \be
    \eta=\left(\begin{array}{cc}
    a & \xi\\
    \xi^* & b\end{array}\right),~~~~~~~~~~a,b\in\R,~~\xi\in\C,~~ab\neq|\xi|^2,
    \label{eta}
    \end{equation}
of the Hermitian matrix $\eta$ and the fact that $D$ is
$\eta$-pseudo-Hermitian for some $\eta$ of the form (\ref{eta}),
i.e., $D^\dagger=\eta D^{-1}\eta^{-1}$ or $D^\dagger\eta D=\eta$,
we find that for $\xi=0$: $r=1$ and $D=D_1$, and for $\xi\neq 0$:
$e^{i\varphi}=e^{i\theta}$ and $D=D_2$. Next, consider the case
that $U$ is not diagonalizable. Then $D$ has the form (\ref{d=}).
Again because $D$ is pseudo-unitary, $\det D\in U(1)$. This
implies $u\in U(1)$, i.e., $u=e^{i\theta}$ for some $\theta\in\R$.
Substituting this expression and the general form (\ref{eta}) of
$\eta$ in $D^\dagger\eta D=\eta$, we find that this equation can
always be satisfied without restricting $\theta$. Therefore, in
this case $D=D_3$.~~~$\square$
    \end{itemize}
The above analysis also yields the form of $\eta$ for each of the
cases considered:
    \begin{itemize}
    \item[1)] For $D=D_1$, there are two possibilities.
        \begin{itemize}
        \item[1.a)] $e^{i\varphi}\neq e^{2i\theta}$: In this case,
        $\xi=0$ and $\eta$ has the diagonal form
            \be
            \eta=\eta_1:=\left( \begin{array}{cc}
            a & 0\\
            0 & b\end{array}\right),~~~~~~~~~~~a,b\in\R-\{0\}.
            \label{eta1a}
            \end{equation}
        Because $a$ and $b$ may have arbitrary sign, the group
        ${\cal U}_{\eta_1}(\C^2)$ is isomorphic to either $U(2)$ or
        $U(1,1)$.
        \item[1.b)] $e^{i\varphi}=e^{2i\theta}$: In this case,
        $D=e^{i\theta} I$ where $I$ is the $2\times 2$ unit matrix.
        Hence, there is no restriction on $\eta$; it has the
        general form (\ref{eta}), and ${\cal U}_{\eta}(\C^2)$ is
        isomorphic to either $U(2)$ or $U(1,1)$.
        \end{itemize}
    \item[2)] If $D=D_2$ with $r=1$ we recover the case 1.b. If
    $D=D_2$ and $r\neq 1$, then $a=b=0$ and $\eta$ has the
    off-diagonal form
        \be
        \eta=\eta_2:=\left(\begin{array}{cc}
        0 & \xi\\
        \xi^* & 0\end{array}\right),~~~~~~~~~~~\xi\in\C-\{0\}.
        \label{eta2}
        \end{equation}
    Because $\eta_2$ is an indefinite matrix,
    ${\cal U}_{\eta_2}(\C^2)$ is isomorphic to $U(1,1)$.
    \item[3)] If $D=D_3$. Then $\eta$ has the general form
        \be
        \eta=\eta_3:=\left(\begin{array}{cc}
        0 & \pm i r e^{-i\theta}\\
        \mp i r e^{i\theta}& 0\end{array}
        \right),~~~~~~~~~~~r\in\R^+,~~\theta\in\R,
        \label{eta3}
        \end{equation}
    and ${\cal U}_{\eta_3}(\C^2)$ is isomorphic to $U(1,1)$.
    \end{itemize}
We can check that the above expressions for $\eta$ are consistent
with the general form of $\eta$ as given in the proof of
Theorem~2. Furthermore, we can obtain the explicit form of the
operator $H:=-i\ln U$. In view of the identity $U=A^{-1}DA$, it is
not difficult to see that if we obtain an operator $H_D$
satisfying $D=e^{iH_D}$, then $H=A^{-1}H_D A$ will satisfy
$U=e^{iH}$. The following table gives the operators $H_D$ and
$\eta$ for $D=D_1,D_2,D_3$. Note that in this table
$\theta,\varphi\in\R$, $r\in\R^+$, $a,b\in\R-\{0\}$,
$\xi\in\C-\{0\}$, and that the trivial case where $D$ is
proportional to the unit matrix is omitted. \vspace{3mm}
\[\mbox{
\begin{tabular}{|c|c|c|c|}
  \hline
  $i$ & $D_i$ & $H_{D_i}$ & $\eta_i$ \\
  \hline
    & & & \\
  $1$ & $\left(\begin{array}{cc}
                e^{i\theta}& 0\\
                0 & e^{i(\varphi-\theta)}\end{array}\right)$
        &  $\left(\begin{array}{cc}
        \theta & 0\\
        0 & \varphi-\theta\end{array}\right)$
            & $\left( \begin{array}{cc}
            a & 0\\
            0 & b\end{array}\right)$ \\
    & & & \\
  \hline
  & & & \\
  $2$ & $\left(\begin{array}{cc}
                r e^{i\theta}& 0\\
                0 & e^{i\theta}/r\end{array}\right)$
    & $\left(\begin{array}{cc}
        \theta-i\ln r & 0\\
        0 & \theta+i\ln r\end{array}\right)$
            & $\left(\begin{array}{cc}
        0 & \xi\\
        \xi^* & 0\end{array}\right)$ \\
& & & \\
  \hline
  & & & \\
  $3$ & $\left(\begin{array}{cc}
            e^{i\theta}& 1\\
            0 & e^{i\theta}\end{array}\right)$
        & $\left(\begin{array}{cc}
        \theta & \theta(e^{i\theta}-1)^{-1}\\
        0 & \theta\end{array}\right)$
            & $\left(\begin{array}{cc}
        0 & \pm i r e^{-i\theta}\\
        \mp i r e^{i\theta}& 0\end{array}
        \right)$ \\
  & & & \\
  \hline
\end{tabular}}
\]

\section{Pseudo-Unitary Dynamical Groups and the Harmonic
Oscillator}

Suppose that $H$ is a $2\times 2$ pseudo-Hermitian matrix serving
as the (time-independent) Hamiltonian for a quantum system,
$U(t):=e^{-it H}$ is the corresponding evolution operator, ${\cal
E}_H$ is the set of all invertible Hermitian $2\times 2$ matrices
$\eta$ satisfying (\ref{ph}), and
    \[ {\cal U}_H:=\bigcup_{\eta\in{\cal E}_H}
    {\cal U}_\eta(\C^2),~~~~~~
       {\cal GU}_H:=\bigcup_{\eta\in{\cal E}_H}
    {\cal GU}_\eta(\C^2),\]
where ${\cal GU}_\eta(\C^2)$ denotes the Lie algebra of ${\cal
U}_\eta(\C^2)$. Then clearly $iH\in {\cal GU}_H$ and for all
$t\in\R$ $U(t)\in {\cal U}_H$. This in particular means that for
each $\eta\in{\cal E}_H$, ${\cal U}_\eta(\C^2)$ serves as a
dynamical group for the quantum system, \cite{nova}. If $H$ is
diagonalizable with a real spectrum then the dynamical group may
be taken to be (isomorphic to) either $U(2)$ or $U(1,1)$ (or one
of their subgroups). If $H$ has (nonreal) complex eigenvalues or
if it is not diagonalizable, then the dynamical group is
necessarily (isomorphic to a subgroup of) $U(1,1)$.\footnote{The
generalization of this statement to arbitrary block-diagonalizable
pseudo-Hermitian Hamiltonians with finite-dimensional blocks is
immediate. If the Hamiltonian is not diagonalizable or has complex
eigenvalues, then the dynamical groups that the system admits are
necessarily (isomorphic to a subgroup of) $U(p,q)$ with $p\neq
0\neq q$.}

A concrete example is provided by the classical equation of motion
for a simple harmonic oscillator of frequency $\omega$,
    \be
    \ddot x+\omega^2 x=0,
    \label{osc}
    \end{equation}
As explained in Refs.~\cite{p4,cqg}, this equation is equivalent
to the Schr\"odinger equation,
    \be
    i\hbar\frac{d}{dt}\Psi=H\Psi,
    \label{sch-eq}
    \end{equation}
where
    \be
    \Psi=:\left(\begin{array}{c}
    x+i\lambda \dot x\\
    x-i\lambda \dot x\end{array}\right),~~~~~~~~
    H=:\frac{\hbar}{2}\left(\begin{array}{cc}
    \lambda\omega^2+\lambda^{-1}&  \lambda\omega^2-\lambda^{-1}\\
    -\lambda\omega^2+\lambda^{-1}& -\lambda\omega^2-\lambda^{-1}
    \end{array}\right),
    \label{osc-H}
    \end{equation}
and $\lambda\in\R^+$ is a time scale. Clearly $H$ is a traceless
matrix. It is also easy to check that $\det H\in\R$ if and only if
$\omega^2\in\R$. Therefore, according to Theorem~3 of
Ref.~\cite{p6}, $H$ is a pseudo-Hermitian matrix provided that
$\omega^2\in\R$. Furthermore, $H$ is diagonalizable unless
$\omega=0$.

In the following we shall only consider the case $\omega^2\in\R$.

For $\omega\neq 0$, we can easily solve the eigenvalue problem and
diagonalize $H$. The corresponding diagonal matrix has the form
$H_D=\hbar\omega\sigma_3$ where $\sigma_3$ is the diagonal Pauli
matrix ${\rm diag}(1,-1)$. Comparing the expression for $H_D$ with
the results given in the above table, we see that $H_D$ is
$\eta$-pseudo-Hermitian with respect to a diagonal metric operator
$\eta$ of the form~(\ref{eta1a}) provided that $\omega^2>0$. In
this case the system admits both the dynamical groups $U(2)$ and
$U(1,1)$. If $\omega^2<0$, $H$ is $\eta$-pseudo-Hermitian with
respect to an off-diagonal metric operator $\eta$ of the
form~(\ref{eta2}) and the system only admits the dynamical group
$U(1,1)$. Finally for $\omega=0$, $H$ is not diagonalizable;
$U=e^{iH}$ has the Jordan canonical form $D_3$; it is
$\eta$-pseudo-unitary for a metric operator $\eta$ of the form
(\ref{eta3}) and the system admits the dynamical group
$U(1,1)$.\footnote{It is interesting to observe that the
noncompact dynamical group $U(1,1)$ arises for the case that
$\omega^2<0$ where Eq.~(\ref{osc}) admits unbounded solutions.}

For the case $\omega^2>0$, the freedom in the choice of the
dynamical group is equivalent to the choice of a positive-definite
or an indefinite inner product on the space of solutions of the
equation~(\ref{osc}), \cite{cqg}. This freedom does not exist if
$\omega^2\leq 0$.

Now, consider changing the parameter $\omega^2$ from a positive
value down to a negative value. If one adopts an indefinite (but
possibly $\omega^2$-dependent) inner product, one can keep $H$
Hermitian with respect to this inner product and view the
evolution operator as tracing a curve in the dynamical group
$U(1,1)$. The best-known example is the Klein-Gordon inner product
that corresponds to the choice $\eta=\sigma_3$, and therefore is
independent of the value of the parameter $\omega^2$. However, if
one initially adopts a (possibly $\omega^2$-dependent)
positive-definite inner product, one cannot maintain the
Hermiticity of $H$ with respect to this inner product once
$\omega^2$ crosses zero. The dynamical group undergoes an abrupt
transition from the group $U(2)$ to the group $U(1,1)$. This
transition may be identified with the change of the signature of
the metric (operator).

For $\omega^2>0$, one may endow the Hilbert space $(\C^2)$ with a
positive-definite invariant inner product. In this case the system
has a $U(2)$ dynamical group and is physically equivalent to the
two-level spin system \cite{nova}, i.e., a spin $1/2$ particle
interacting with a fixed magnetic field. The time-evolution is
clearly unitary. This equivalence is destroyed once $\omega^2$
becomes nonpositive. In this case the dynamical group is $U(1,1)$
and the system does not admit a unitary evolution with respect to
any positive-definite inner product on $\C^2$. For the case that
$\omega^2>0$ one could as well choose an indefinite invariant
inner product (this is precisely what was done historically). But
such a choice leads to a non-unitary quantum system with a
two-dimensional Hilbert space and a $U(1,1)$ dynamical group. As
is well-known the corresponding quantum harmonic oscillator also
has a $U(1,1)$ (or rather $SU(1,1)$) dynamical group, \cite{nova}.
Therefore, as far as the dynamics is concerned the non-unitary
system describing the dynamics of the classical oscillator is
equivalent to the unitary quantum harmonic oscillator.

For the case $\omega^2>0$ there are therefore two alternatives.
One is to choose a positive-definite invariant inner product which
corresponds to the dynamical group $U(2)$. The other is to choose
an indefinite invariant inner product which leads to the dynamical
group $U(1,1)$.

Now, suppose that one wishes to keep the same dynamical group but
insists on being able to describe the dynamics using a unitary
quantum system. In the first alternative this is already the case.
But in the second alternative one needs to use an
infinite-dimensional Hilbert space, because being a noncompact Lie
group $U(1,1)$ does not admit a finite-dimensional unitary
representation. Therefore {\em it is the demand for unitarity that
leads to the quantization of the oscillator.} The latter is
however not unique because $U(1,1)$ has inequivalent unitary
irreducible representations. This does not lead to any problems,
because the dynamics always takes place in the dynamical group
\cite{nova}. As a result the dynamical aspects of all possible
quantum systems associated with the classical harmonic oscillator
are equivalent.\footnote{Note that here `quantization' does not
means the canonical quantization which is unique in the sense that
the Weyl-Heisenberg algebra has a unique irreducible (projective)
representation. It means defining the Hilbert space as the
representation space of a unitary irreducible projective
representation of the dynamical group, and representing the
Hamiltonian as an element of the Lie algebra of the dynamical
group.}

The above two alternative are also available in describing free
Klein-Gordon fields (or more generally Klein-Gordon fields
interacting with a stationary magnetic field). The second
alternative applies more generally even to the cases of
interacting fields. It corresponds to Dirac's method of second
quantization that forms the foundations of quantum field theories.
The first alternative was noticed quite recently,
\cite{cqg,ap,p57}.\footnote{See however Refs.~\cite{old} that were
brought to the author's attention after the completion of this
project.} Its advantage is to provide a genuine probability
interpretation for first quantized Klein-Gordon fields \cite{p59}.
Its main application is in quantum cosmology \cite{ap}.

\section{Conclusion}

In this article, we explored various properties of pseudo-unitary
operators and proved a spectral characterization theorem for the
class of block-diagonalizable pseudo-unitary operators with
finite-dimensional blocks. We applied our results to clarify the
structure of pseudo-unitary matrices paying attention to the role
of the inner product and the fact that it is not unique. We showed
that the relationship between Hermitian and unitary matrices
generalize to the pseudo-Hermitian and pseudo-unitary matrices.
Specifically every pseudo-unitary matrix is the exponential of $i$
times a pseudo-Hermitian matrix.

We showed that the symplectic transformations of classical
mechanics are certain pseudo-unitary and pseudo-Hermitian
operators. This led to a proof of the spectral theorem for
symplectic matrices and to the identification of the symplectic
groups $Sp(2n)$ with the real subgroups of a matrix group that is
isomorphic to the pseudo-unitary groups $U(n,n)$. The description
of the symplectic transformations in terms of pseudo-unitary and
pseudo-Hermitian operators suggests the possibility of the
application of the latter in classical mechanics.\footnote{For a
related discussion see \cite{cm}.}

Furthermore, we derived the canonical forms of arbitrary $2\times
2$ pseudo-unitary matrices, and studied the pseudo-unitary system
describing a classical harmonic oscillator. For real nonzero
frequencies, this system admits both the dynamical groups $U(2)$
or $U(1,1)$. If one imposes the condition of the unitarity of the
evolution, then the choice $U(2)$ identifies the dynamics of the
oscillator with that of a two-level quantum system, and the choice
$U(1,1)$ leads to a quantization of the oscillator. This picture
provides a rather interesting link between the demand for
unitarity and the need for (second) quantization.

\section*{Acknowledgment}

This work has been supported by the Turkish Academy of Sciences in
the framework of the Young Researcher Award Program
(EA-T$\ddot{\rm U}$BA-GEB$\dot{\rm I}$P/2001-1-1). The author
acknowledges the constructive comments of the referee that led him
to explore the connection with symplectic transformations.

\ed